\documentclass[11pt,twoside]{article}
\usepackage{asp2014}

\aspSuppressVolSlug
\resetcounters

\begin{document}
\title{Characterising radio telescope software with the Workload Characterisation Framework}

\author{Y.~G.~Grange$^1$, R.~Lakhoo$^2$\textsuperscript{*}, M.~Petschow$^1$, C.~Wu$^3$, B.~Veenboer$^1$, I.~Emsley$^2$, T.~J.~Dijkema$^1$, A.~P.~Mechev$^4$, and G.~Mariani$^5$
	\affil{$^1$ ASTRON, The Netherlands; \email{(grange@astron.nl)}}
	\affil{$^2$ Oxford e-Research Centre, University of Oxford, United Kingdom}
	\affil{$^3$ ICRAR, University of Western Australia, Australia}
	\affil{$^4$ Leiden Observatory and LIACS, Leiden University, The Netherlands}
	\affil{$^5$IBM Research, The Netherlands}
	}
	
\paperauthor{Y.G. Grange}{grange@astron.nl}{0000-0001-5125-9539}{ASTRON}{R\&D}{Dwingeloo}{}{7991PD}{The Netherlands}
\paperauthor{R. Lakhoo}{rahim.lakhoo@oerc.ox.ac.uk}{0000-0003-4624-5639}{University of Oxford}{Oxford e-Research Centre}{Oxford}{}{}{United Kingdom}
\paperauthor{M. Petschow}{petschow@astron.nl}{}{ASTRON}{R\&D}{Dwingeloo}{}{7991PD}{The Netherlands}
\paperauthor{C. Wu}{chen.wu@uwa.edu.au}{}{University of Western Australia}{International Centre for Research in Radio Astronomy}{Perth}{}{}{Australia}
\paperauthor{B. Veenboer}{veenboer@astron.nl}{}{ASTRON}{R\&D}{Dwingeloo}{}{7991PD}{The Netherlands}
\paperauthor{I. Emsley}{Iain.emsley@oerc.ox.ac.uk}{0000-0002-5805-1367}{University of Oxford}{Oxford e-Research Centre}{Oxford}{}{}{United Kingdom}
\paperauthor{T.J. Dijkema}{dijkema@astron.nl}{}{ASTRON}{R\&D}{Dwingeloo}{}{7991PD}{The Netherlands}
\paperauthor{A.P. Mechev}{apmechev@strw.leidenuniv.nl}{}{Universiteit Leiden}{Leiden Observatory}{Leiden}{}{}{The Netherlands}
\paperauthor{G. Mariani}{}{giovanni.mariani@nl.ibm.com}{IBM Research}{}{}{}{}{The Netherlands}
\begin{abstract}
	We present a modular framework, the {\em Workload Characterisation Framework}
	(WCF), that is developed to reproducibly obtain, store and compare key
	characteristics of radio astronomy processing software. As a demonstration, we
	discuss the experiences using the framework to characterise a LOFAR calibration
	and imaging pipeline.
\end{abstract}
\section{Introduction}
Modern low-frequency radio interferometers, such as
the LOw Frequency ARray (LOFAR)~\citep{vanhaarlem2013}, consist of many antennas and data processing is performed by software; this is reflected by the community term
{\em software telescope}. Such software telescopes are modular and under constant
development by large interdisciplinary research collaborations, working across
continents; consequently, obtaining a holistic view of the entire system becomes
a challenging task. It is a considerable challenge to manage, process and store
such large datasets, within budgets and constraints, while ``pushing the
envelope'' of technology. To achieve such cost-performance optimum, the first aim
is  to  obtain  a  quantitative  understanding  of  the compute, energy and data
access  behaviours  exhibited by various radio astronomy data processing
software pipelines and algorithms. Using
low-level Linux kernel and hardware interfaces, the {\em Workload
	Characterisation Framework} (WCF) can be used to support operations and
software development.

To encourage best practices, the work undertaken requires to be reproducible:
the software should be open and publicly available, and it should allow for
others to reproduce the software environment.  In this work, we present the WCF
that aims to provide key metrics to characterise workloads in a standard and
reproducible manner. The WCF is designed to be extensible and we present an
example of such an extension that aims to provide further insight when
identifying software bottlenecks for research and development
(R\&D) purposes.
\section{The Workload Characterisation framework}

The WCF is under development as part of the Local Monitoring and Control (LMC)
work package of the SKA Science Data Processor (SDP) consortium. The first
purpose of the WCF is to provide the SDP compute resource scheduler with the
essential workload characteristics for each pipeline processing component, in
order to make optimal (cost-effective) scheduling decisions. Secondly, the WCF
aims to assist telescope developers and operators to obtain a quantitative
understanding of the compute, energy and data access behaviours exhibited by
various pipelines, components and algorithms.  Finally, the WCF aims to enable
the micro-benchmarking of different compute platforms, which can be used for
optimisation and comparison.

The WCF prototypes have already been used, or tested for SKA precursor and
pathfinders, including LOFAR, MWA, JVLA HDR, ASKAPSoft and the CHILIES project.
\section{The LOFAR use case}
The WCF can be used for a variety of tasks; we focus on a bottleneck analysis --
with the aim to support more targeted system development efforts (software and
hardware).  In addition to the typical WCF output data to assist this task, we
consider non-time-series data to create the link between system behaviour and
the structure of the software.

In this section, as an example use case, we consider a calibration and imaging
processing pipeline, hereafter referred to as {\sc Calib}, which creates sky
images from the LOFAR telescope.

\subsection{The {\sc Calib} pipeline}
The {\sc Calib} pipeline has been used to image HBA commissioning data of the
Galactic diffuse synchrotron emission~\citep{iacobelli2013}.

\begin{figure}[bht]
	\centering
	\includegraphics[width=0.9\textwidth, trim={50 300 20 75}, clip]{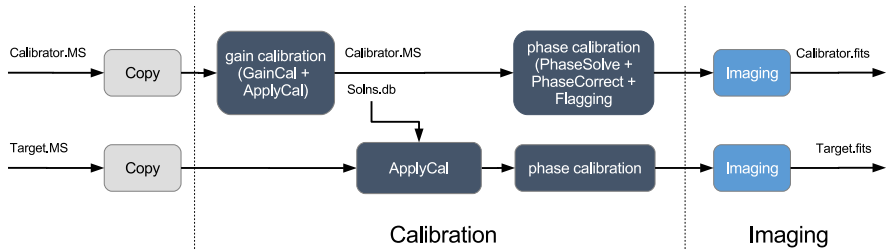}
	\caption{Work flow of the {\sc Calib} pipeline.}
	\label{fig:flow}
\end{figure}

For each frequency band, there are two pointings: one towards the target field
and the other towards the calibrator field. As depicted in
Fig.~\ref{fig:flow}, the two input measurement sets are fed into the pipeline.
After a copy operation, the calibrator is used to solve for the antenna gain
amplitudes, which are then applied to the target.  A phase calibration is then
performed on both the calibrator and the target data. Finally, sky images are
derived from the calibrated data sets.  The entire calibration is performed
using LOFAR's DPPP (data preprocessing pipeline); the imaging is performed using
WSClean~\citep{offringa2014}.

\subsection{Goal-oriented system development \label{sec:goals}}

With goal-oriented or targeted system development, we promote the following work
flow: (1) for a given micro-benchmark and platform, evaluate the baseline
performance of both the hardware and software -- ideally by replicating the
results of measurements stored in a data base; (2) use the WCF to analyse system
behaviour and identify bottlenecks; (3) concentrating on a specific part and
possibly using other performance analysis tools, optimize the code or change the
underlying hardware; (4) rerun the WCF measurements of the entire pipeline to
assess the impact of the changes and store the results in a database.

With the WCF we assist a goal-oriented development in the following
way:
(1) Evaluating a pipeline's performance over a long time requires a
standardised storage format of the measurements.
(2) By standardising the measurement tools, the measurements become comparable over a
wide range of use cases: comparing different algorithms and their implementations, different
hardware, and different input data and data formats. Among other advantages, this allows
for meta-analysis of stored measurements for various configurations.
(3) Quickly identify computational and other resource
bottlenecks. This information could be used to alleviate software bottlenecks
and identify the most salient system features that determine
performance. Besides facilitating such a hardware-software co-design
procedure, the insights can be used to inspire general research.
\section{Results}

\begin{figure}[bht]
	\begin{center}
		\includegraphics[width=\textwidth,trim=0mm 1mm 0mm 12mm, clip]{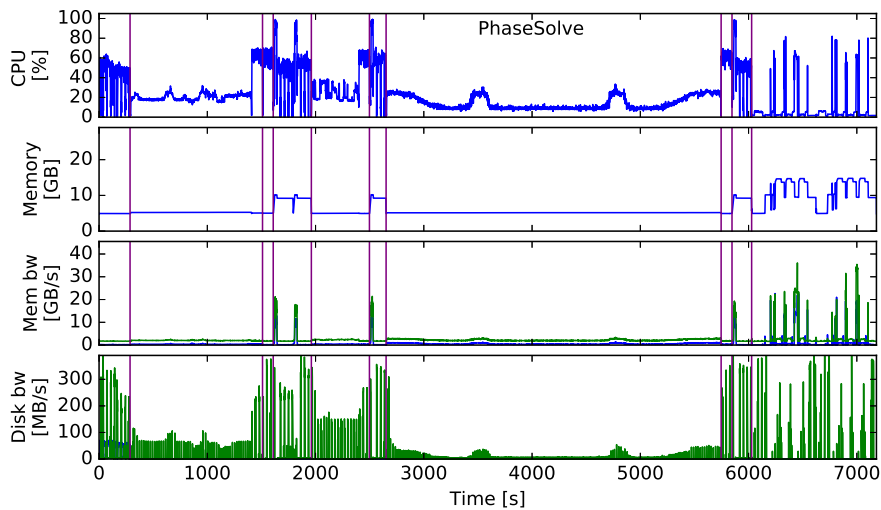}
		\caption{Time series of system characteristics. \label{fig:timeline}}
	\end{center}
\end{figure}

In Fig.~\ref{fig:timeline} we present time series data of a {\sc Calib} pipeline
run. We choose the metrics CPU and memory usage, memory bandwidth, and disk I/O
bandwidth -- all important parameters to identify performance bottlenecks. While
a detailed analysis of the data is beyond the scope of this paper, we
demonstrate its use for one specific step of the pipeline, {\sc PhaseSolve}.
This step requires roughly 55\% of the entire execution time
(Fig.~\ref{fig:sunburst}; left). Further analysis indicates that the step
exhibits poor scaling behaviour: the execution using a 28 core dual-socket Intel
Xeon compute node is only $1.23$ times faster than a sequential execution
(Fig.~\ref{fig:sunburst}; right). {Using the OeRC SKA testbed, low-level CPU
	characteristics were gathered, and showed less than optimal CPU core usage on
	average, a high number of CPU migrations (292/sec) and context switches
	(3,521/sec). The poor scaling characteristics were verified by configuring OMP
	affinity settings, using a Round-robin (RR) real-time CPU scheduler and CPU
	pinning, only resulted in a 8.2\% increase in runtime using 10 fewer CPU cores
	than the default.  A natural next step is to improve on the scalability of the
	{\sc PhaseSolve} step of the pipeline.
	
	\begin{figure}[ht]
		\begin{center}
			\includegraphics[width=0.48\textwidth,trim=50mm 175mm 50mm 26mm, clip]{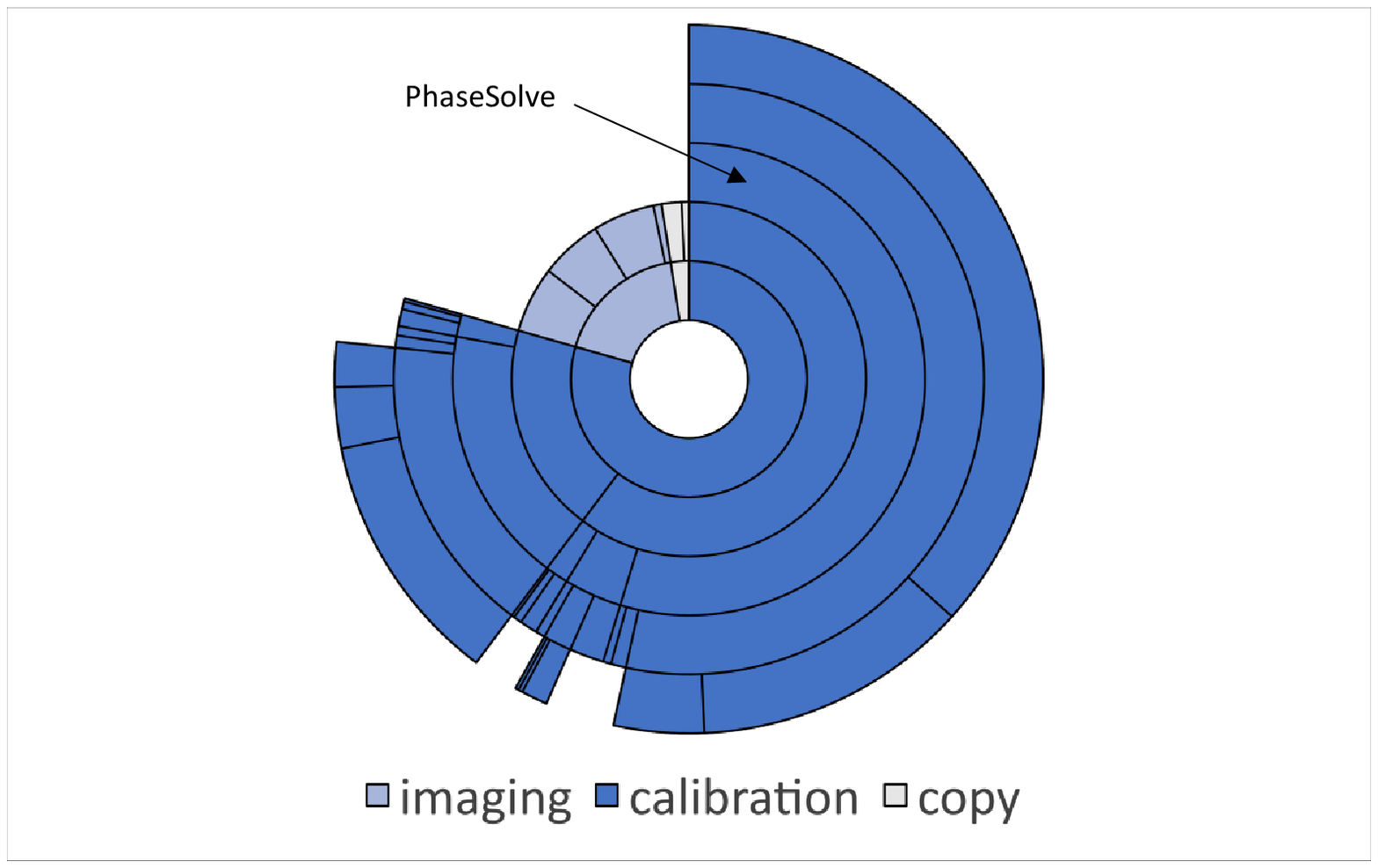}
			\includegraphics[width=0.48\textwidth,trim=1mm 0mm 15mm 10mm, clip]{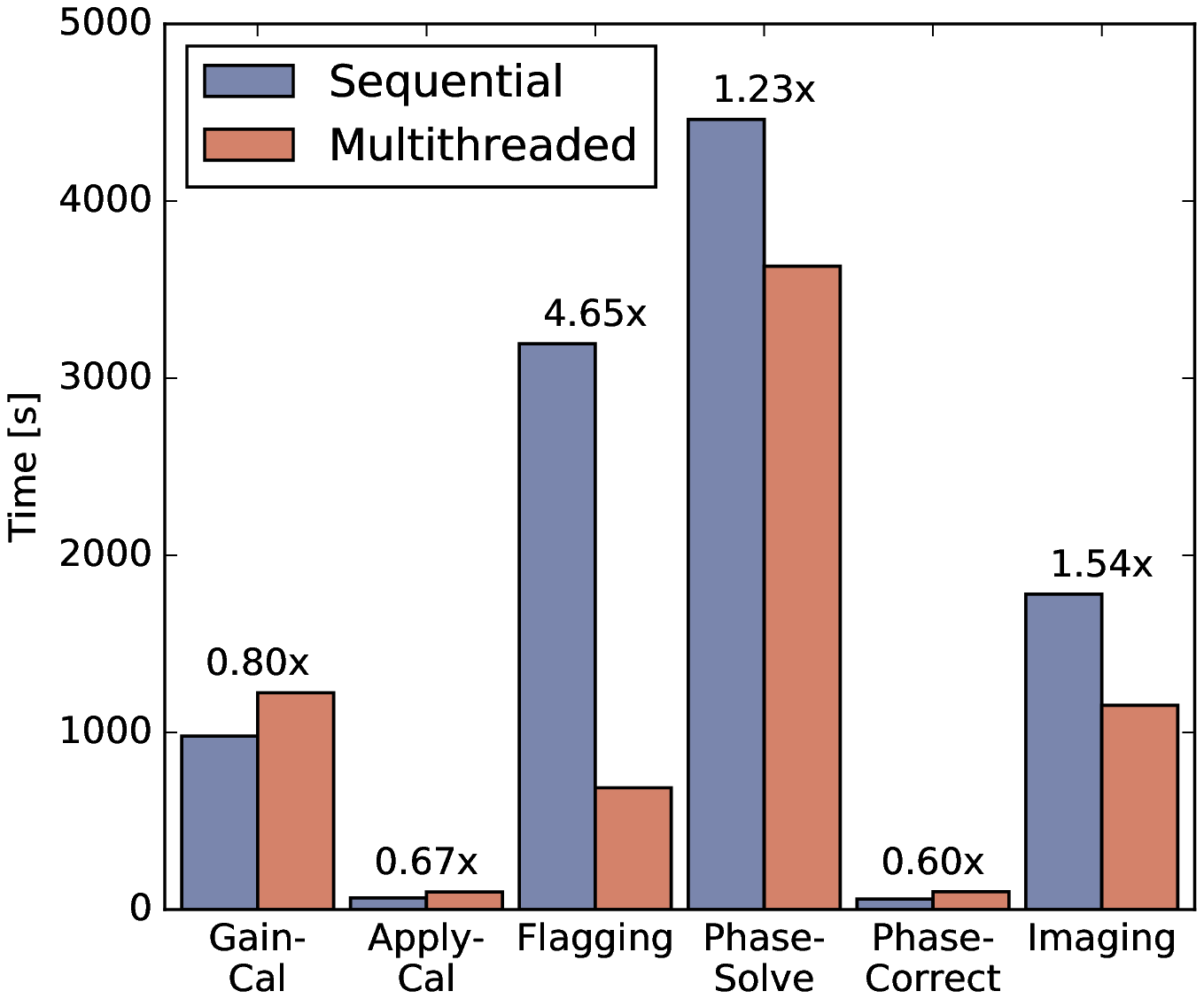}
			\caption{Breakdown of execution time and speedup due due parallel execution.
				\label{fig:sunburst}}
		\end{center}
	\end{figure}
	
	\section{Conclusion}
	Using a LOFAR calibration and imaging pipeline, we demonstrated the use of the
	WCF for R\&D radio telescope software and hardware development. Such structured
	approach is especially relevant for the development of new instruments and
	algorithms for the SKA.
\acknowledgements{
	We thank Marco Iacobelli for providing the {\sc Calib} processing pipeline and
	the input data.  This work is supported by the Dutch Ministry of EZ and the
	province of Drenthe through the ASTRON-IBM Dome grant and the EU FP7 under grant
	no ICT-610476 (DEEP-ER). The authors acknowledge the Department of Physics and
	the OeRC at the University of Oxford, for use of their system resources and
	personnel. The authors would like to acknowledge the use of the University of
	Oxford Advanced Research Computing facility in carrying out this work.
	}

\bibliography{P8-14}

\end{document}